\documentclass[12pt,showpacs,superscriptaddress,prl,floatfix]{revtex4}
\usepackage{graphicx}
\usepackage{epsfig}
\usepackage[usenames]{color}
\usepackage{subeqn}
\usepackage{ulem} 

\newcommand{\bk}{{\bf k}}
\newcommand{\br}{{\bf r}}
\newcommand{\bt}{{\bf t}}
\newcommand{\bu}{{\bf u}}
\newcommand{\bx}{{\bf x}}
\newcommand{\by}{{\bf y}}
\newcommand{\bz}{{\bf z}}

\newcommand{\bS}{{\bf S}}
\newcommand{\bT}{{\bf T}}

\newcommand{\kb}{{k_{\rm B}}}
\newcommand{\kbt}{{k_{\rm B}T}}

\date{\today}

\begin{document}
\title{Anisotropic interaction of two-level systems with acoustic waves in disordered crystals}

\author{Drago\c s-Victor Anghel}
\affiliation{Department of Theoretical Physics, National Institute for Physics and Nuclear Engineering--''Horia Hulubei'', Str. Atomistilor no.407, P.O.BOX MG-6, Bucharest - Magurele, Romania}

\author{Dmitry Churochkin}
\affiliation{Saratov State University, 410012, Astrakhanskaya St. 83, Saratov, Russia}

\begin{abstract}
We apply the model introduced in Phys. Rev. B {\bf 75}, 064202
(2007), cond-mat/0610469, to calculate 
the anisotropy effect in the interaction of two level systems with 
phonons in disordered crystals.
We particularize our calculations to cubic crystals
and compare them with the available experimental data to extract
the parameters of the model. With these parameters we calculate the 
interaction of the dynamical defects in the disordered crystal with phonons 
(or sound waves) propagating along other crystalographic directions, providing 
in this way a method to investigate if the anisotropy comes from the
two-level systems being preferably oriented in a certain
direction or solely from the lattice anisotropy with the
two-level systems being isotropically oriented.
\end{abstract}

\maketitle

\section{Introduction} \label{intro}

The low temperature acoustic and thermal properties of amorphous,
glassy materials are remarkably
similar and they can be explained to a large extent by assuming that
the material contains a large number of dynamic defects. These dynamic
defects are tunneling systems (TS) and are modeled by an ensemble of
two-level systems (TLS)
\cite{JLowTempPhys.7.351.1972.Philips,PhilMag.25.1.1972.Anderson}.
Crystals with defects--with a large enough amount of
disorder--exhibit also glass-like properties, but these properties
are not so universal and, even more, they are not
isotropic--like it is, for example, the sound absorbtion and
velocity change, which depends on the crystalographic direction in
which the sound propagates \cite{PhysRevB.51.8158.1995.Laermans}.

Since a detailed microscopic model of tunneling systems in glassy
materials is still not available, the study of disordered crystals
is especially interesting because it offers an additional
opportunity for their clarification: in some materials we know
quite well which are the entities that tunnel between different
equilibrium positions. Beside this, the anisotropy of the
TLS-sound wave interaction in crystals represents another
challange to the interaction models of TLSs which requires
clarification.

In this paper we give an explanation for the anisotropy observed
in the glass-like properties of general, disordered crystals, by
employing a model recently published
\cite{PhysRevB75.064202.2007.Anghel}. In this model we assume that
each TLS is characterized by a direction in space, call it
$\hat\bt$--this might be the direction defined by the two
potential wells of the tunneling system, or the axis of rotation
of the tunneling entity--and we introduce a coupling between the
TLS and a strain field $[S]$, which is dependent on the amplitude
of $[S]$ at the position of the TLS and on the orientation of
$\hat\bt$ with respect to $[S]$. In \cite{PhysRevB75.064202.2007.Anghel} 
the model was applied to an amorphous solid, assuming that the 
directions $\hat\bt$ are isotropically distributed, and the
effective coupling of an elastic wave with a TLS was calculated as the average
over the directions of the TLS. 
In this way it was proven on very general grounds that, on
average, the longitudinal waves couple with the TLSs stronger than
the transversal waves--in standard notations,
$\gamma_l>\gamma_t$ \cite{PhysRevB75.064202.2007.Anghel}.

In a disordered crystal with TLSs, there could be at least two
sources of anisotropy. The first one is that the TLSs might not be
anymore isotropically oriented so the effective coupling of
elastic waves with them depends, through $\hat\bt$, on the waves
direction of propagation and on their polarization. The
second source of anisotropy is that besides the relative
orientation of $\hat\bt$ and $[S]$, the symmetry of the crystal is
manifested also in the interaction of elastic waves with TLSs
\cite{PhysRevB75.064202.2007.Anghel,PhysRevB76.165425.2007.Kuhn}.
This leads to anisotropy effects in the interaction of elastic
waves with TLS even if the TLS distribution is isotropic.

In this paper we shall analyse mainly the second type of
anisotropy. We shall assume that the TLS orientations are
isotropically distributed and we shall calculate the anisotropy
effects imposed only by the lattice symmetries onto the
interaction hamiltonian. Existing experimental data (see Ref.
\cite{Topp:thesis} and references therein) will enable us to
obtain relations between the parameters of the model for a cubic
lattice. Using these parameters we can make predictions about the
absorbtion of elastic waves propagating along other
crystalographic directions or having different polarizations.
These predictions could constitute a first test for the isotropy
of the TLS orientations in a specific crystal.

\section{Anisotropic interaction of two-level systems with sound waves}

In the \textit{standard tunneling model} (STM), the Hamiltonian of an
isolated TLS is written in a two-dimensional basis as
\cite{JLowTempPhys.7.351.1972.Philips,PhilMag.25.1.1972.Anderson}
\begin{equation}
  \label{eqn_TLS_hamiltonian}
  H_{\text{TLS}}
      =  \frac{\Delta}{2}\sigma_z -\frac{\Lambda}{2}\sigma_x
 \equiv \frac{1}{2}\left(\begin{array}{cc}
                     \Delta   & -\Lambda\\
                     -\Lambda & -\Delta \end{array}
                \right)
\end{equation}
where $\Delta$ is called the {\em asymmetry of the potential}
and $\Lambda$ the {\em tunnel splitting}. The basis in which is
written the Hamiltonian of Eq. (\ref{eqn_TLS_hamiltonian}) is chosen
in such a way that a perturbation to the TLS, caused by a
strain field, say $[S]$, is described by a diagonal Hamiltonian,
\begin{eqnarray}
H_1
 &=& \frac{1}{2}\left(\begin{array}{cc}
                \delta&0\\
                0&-\delta\end{array}\right)
                \,.\label{eqn_H_1_prime}
\end{eqnarray}
with $\delta\equiv2[\gamma]:[S]$ and $[\gamma]$ a second rank
tensor of coupling constants; by ``$:$'' we denote the dyadic
product. Typically, in the STM one considers the coupling of TLSs
with transversally or longitudinally polarized sound waves, so not
too much attention has been given to the $[\gamma]$ tensor and in
general $\delta$ is written simply as
$\delta=2\gamma_{l,t}S_{l,t}$, with $\gamma$ and $S$ being scalars
($S$ is the amplitude of the strain field) and $l$ and $t$
denoting the longitudinal ($l$) or the transversal ($t$)
polarization of the sound wave, respectively. Such a simple
description of the TLS-strain field interaction has several
shortcomings--e.g. $\delta$ is not invariant and even leads to
physical ambiguities at the rotation of the coordinates axes--and
cannot account for the different coupling of the TLS with sound
waves propagating in different directions. In consequence, in Ref.
\cite{PhysRevB75.064202.2007.Anghel} it is proposed a model which
eliminates the shortcomings and takes into account the symmetries
of the material in which the TLSs are embeded and the orientation
of the TLS with respect to the strain field. Let us describe
briefly how this is done.

We construct form the components of $\hat\bt$ the simple
$3\times3$ symmetric tensor, $[T]$, of components $T_{ij}=t_it_j$
and we introduce the forth rank tensor of TLS-strain field
coupling constants, $[[R]]$. With these two objects, we build the
general tensor $[\gamma]$, as
$\gamma_{ij}=T_{kl}R_{klij}$--throughout this paper we assume
\textit{summation over the repeated indices}. The forth rank
tensor $[[R]]$ has a similar structure as the forth rank tensor
$[[c]]$ of stifness constants and reflects the symmetries of the
crystal that contains the TLS
\cite{PhysRevB75.064202.2007.Anghel,arXiv.org:0710.0720.Anghel}.

For the convenience of the calculations we work here, like in
Refs.
\cite{PhysRevB75.064202.2007.Anghel,PhysRevB76.165425.2007.Kuhn,arXiv.org:0710.0720.Anghel},
in abbreviated subscript notations and write $[T]$ and $[S]$ as
the six elements vectors,
$\bT\equiv(T_{11},T_{22},T_{33},2T_{23},2T_{13},2T_{12})^t$ and
$\bS\equiv(S_{11},S_{22},S_{33},2S_{23},2S_{13},2S_{12})^t$, where
by ``$\cdot^t$'' we denote the transpose. Following the notations
of Auld \cite{Auld:book}, the components of the symmetric tensors
will be denoted in abbreviated subscript notations by a single,
upper case subscript--e.g. $T_I$, $S_I$, and $T_3\equiv
T_{33}=t_3^2$; also in abbreviated subscript notations, the
tensors $[[R]]$ and $[[c]]$ will be written as $6\times6$
matrices, $[R]$ and $[c]$, of components $R_{IJ}$ and $c_{IJ}$,
respectively. Putting all these together we get the expression
$\delta=2\bT^t\cdot[R]\cdot\bS$
\cite{PhysRevB75.064202.2007.Anghel,PhysRevB76.165425.2007.Kuhn,arXiv.org:0710.0720.Anghel}.

Having now the full expression for the interaction Hamiltonian,
$H_1$, we can calculate the amplitude of excitation of a TLS, of
parameters $\Delta$ and $\Lambda$, by a phonon of wave-vector
$\bk$ and polarization $\sigma$; we denote by $n_{\bk\sigma}$ the
number of phonons on the mode $(\bk,\sigma)$ after the TLS
excitation process. The displacement field of the phonon,
$\bu_{\bk\sigma}$, is normalized to
$N_{\bk\sigma}\equiv\sqrt{\hbar/(2V\rho\omega_{\bk\sigma})}$, and
has the strain field $\bS_{\bk\sigma}=\nabla_S\bu_{\bk\sigma}$
(where by $\nabla_S$ we denote the \textit{symmetric gradient}).
This way we get
\begin{equation} \langle
n_{\bk\sigma},\uparrow|\tilde{H}_1|n_{\bk\sigma}+1,\downarrow\rangle_{\bu_{\bk\sigma}}
 = -\frac{\Lambda}{\epsilon}\sqrt{n_{\bk\sigma}}\bT^t\cdot[R]\cdot\bS_{\bk\sigma}
\label{eqn_matrix_element}
\end{equation}
where $\epsilon=\sqrt{\Delta^2+\Lambda^2}$ is the
\textit{excitation energy of the TLS}. Therefore the
\textit{phonon scattering rate by a TLS in the ground state} is
\begin{equation}
\Gamma_{\bk\sigma}(\hat\bt) = \frac{2\pi}{\hbar}\frac{\Lambda^2n_{\bk\sigma}}
{\epsilon^2}|\bT^t\cdot[R]\cdot\bS_{\bk\sigma}|^2\delta(\epsilon-\hbar\omega).
\label{eqn_Gamma_bar}
\end{equation}
The main characteristic of the TLS-elastic strain interaction,
is contained in the quantity
$M_{\bk,\sigma}(\hat\bt)\equiv\bT^t\cdot[R]\cdot\bS_{\bk\sigma}$, which
we shall calculate next.

As mentioned above, $[R]$, like $[c]$, reflects the symmetries of the lattice.
The most general type of lattice is triclinic, in
which case $[R]$ is symmetric and contains 21 independent constants.
Such a lattice is very complex and in general does not sustain
simple transversally or longitudinally polarized elastic waves, but
instead, the elastic waves propagating through the crystal will be
complex superpositions of longitudinally and transversally polarized
plane-waves.
So, to start with a simpler case and also to be able to compare
our calculations with available experimental data
\cite{Topp:thesis}, we shall focus in this paper on lattices
with cubic symmetry. The tensor $[R]$ for the cubic lattice is
very similar to the one for an isotropic material
\cite{PhysRevB75.064202.2007.Anghel,PhysRevB76.165425.2007.Kuhn,arXiv.org:0710.0720.Anghel},
but it contains 3 independent constants instead of 2, like in the
isotropic case. So we can preserve the notations of Refs.
\cite{PhysRevB75.064202.2007.Anghel,PhysRevB76.165425.2007.Kuhn,arXiv.org:0710.0720.Anghel},
and write
\begin{eqnarray}
[R] &=& \tilde\gamma\cdot \left(\begin{array}{cccccc}
1&\zeta&\zeta&0&0&0\\
\zeta&1&\zeta&0&0&0\\
\zeta&\zeta&1&0&0&0\\
0&0&0&\xi&0&0\\
0&0&0&0&\xi&0\\
0&0&0&0&0&\xi
\end{array}\right), \label{R_cubic}
\end{eqnarray}
without imposing the isotropy constraint, $\zeta+2\xi=1$;
similarly, the tensor of elastic stiffness constants is
\begin{eqnarray}
[c] &=& \left(\begin{array}{cccccc}
c_{11}&c_{12}&c_{12}&0&0&0\\
c_{12}&c_{11}&c_{12}&0&0&0\\
c_{12}&c_{12}&c_{11}&0&0&0\\
0&0&0&c_{44}&0&0\\
0&0&0&0&c_{44}&0\\
0&0&0&0&0&c_{44}
\end{array}\right). \label{c_cubic}
\end{eqnarray}

Using $[c]$ we can write the Christoffel equation to find $\bu$
and $\bS$ for the elastic waves propagating in different
directions and then we can calculate $M$ for any $\hat\bt$,
$\Delta$ and $\Lambda$. In the end, we average over the ensemble
of TLSs, to determine the attenuation of the elastic wave or the
scattering rate of the phonon. We shall apply this procedure for
strain fields corresponding to elastic waves propagating along the
crystalographic directions $\langle100\rangle$,
$\langle110\rangle$ and $\langle111\rangle$ of the cubic lattice.
Along these directions, the cubic lattice can sustain simple,
longitudinally and transversally polarized elastic waves, for any
allowed values of the parameters $c_{11}$, $c_{12}$, and $c_{44}$.

Solving the Christoffel equations we find that the sound velocities of the
longitudinal waves propagating in the $\langle100\rangle$,
$\langle110\rangle$ and $\langle111\rangle$ directions are
$c_{l,\langle100\rangle}=\sqrt{c_{11}/\rho}$,
$c_{l,\langle110\rangle}=\sqrt{(c_{11}+c_{12}+c_{44})/\rho}$, and
$c_{l,\langle111\rangle}=\sqrt{(c_{11}+2c_{12}+2c_{44})/\rho}$, respectively.
Similarly, the sound velocities of the transversal waves propagating
in the $\langle100\rangle$ and $\langle111\rangle$ directions are
$c_{t,\langle100\rangle}=\sqrt{c_{44}/\rho}$ and
$c_{t,\langle111\rangle}=\sqrt{(c_{11}-c_{12}-c_{44})/\rho}$, respectively,
whereas for the transversal waves propagating in the $\langle110\rangle$
direction the sound velocity depends on the direction of polarization:
if the wave is polarized in the $\langle100\rangle$ direction (and
perpendicular on the direction of propagation), the
sound velocity is $c_{t,\langle110\rangle}^{\langle100\rangle}=\sqrt{c_{44}/\rho}$,
and if the wave is polarized in the $\langle110\rangle$ direction 
(and also perpendicular on the direction of propagation), the
sound velocity is
$c_{t,\langle110\rangle}^{\langle110\rangle}=\sqrt{(c_{11}-c_{12}-c_{44})/\rho}$.
Now we can calculate $M$ for these three directions of propagation.

Since the three directions, $\hat\bx$, $\hat\by$, and $\hat\bz$ are
equivalent, let us take the $\langle100\rangle$ direction as the
$\hat\bz$ direction. We also define $\hat\bt$  by the angles
$\theta$ (nutation) and $\phi$ (precession), as
$\hat\bt\equiv(\sin\theta\cos\phi,\sin\theta\sin\phi,\cos\theta)^t$.
With these conventions, we get for the longitudinal wave,
$\bu_{k\hat\bz,l}(\br) = N\hat\bz e^{ik\hat\bz\cdot\br}$,
\begin{subequations}\label{M100}
\begin{equation} \label{M100l}
M_{k\hat\bz,l} = ik\tilde\gamma N_{k\hat\bz,l}[\zeta+\cos^2\theta(1-\xi)]
\end{equation}
(to simplify the expressions without reducing the clarity, we
shall always drop the exponential from the expressions of $M$ and
the subscripts of $N$; the implicit subscripts of $N$ are always the same
as the ones of $M$ and $\bu$) and for the two, reciprocally
perpendicular, transversal waves, $\bu_{k\hat\bz,t,x}(\br) =
N\hat\bx e^{ik\hat\bz\cdot\br}$ and $\bu_{k\hat\bz,t,y}(\br) =
N\hat\by e^{ik\hat\bz\cdot\br}$,
\begin{eqnarray}
M_{k\hat\bz,t,x} &=& ik\tilde\gamma\xi N\sin(2\theta)\cos(\phi)
\label{M100tx} \\
M_{k\hat\bz,t,y} &=& ik\tilde\gamma\xi N\sin(2\theta)\sin(\phi)
\label{M100ty}
\end{eqnarray}
\end{subequations}

For the waves propagating in the $\langle111\rangle$ direction we get the
following results. For the longitudinal wave,
$\bu_{{k(\hat\bx+\hat\by+\hat\bz)}/{\sqrt{3}},l}(\br) = N\frac{\hat\bx+\hat\by+\hat\bz}{\sqrt{3}}\exp\left[ik\frac{\hat\bx+\hat\by+\hat\bz}{\sqrt{3}}\cdot\br\right]$,
\begin{subequations}\label{M111}
\begin{eqnarray}
M_{{k(\hat\bx+\hat\by+\hat\bz)}/{\sqrt{3}},l} &=&
N\frac{ik\tilde\gamma}{3}\{[2\sin(2\theta)(\sin\phi+\cos\phi) \label{M111l} \\
&& +2\sin(2\phi)\sin^2(\theta)]\xi+2\zeta+1\}\nonumber
\end{eqnarray}
and for the two transversal waves,
$\bu_{{k(\hat\bx+\hat\by+\hat\bz)}/{\sqrt{3}},t,p_1}(\br)=N\hat p_1
\exp\left[ik\frac{\hat\bx+\hat\by+\hat\bz}{\sqrt{3}}\cdot\br\right]$
and
$\bu_{{k(\hat\bx+\hat\by+\hat\bz)}/{\sqrt{3}},t,p_2}(\br)=N \hat p_2
\exp\left[ik\frac{\hat\bx+\hat\by+\hat\bz}{\sqrt{3}}\cdot\br\right]$,
with polarizations $\hat p_1=\frac{-\hat\bx+\hat\bz}{\sqrt{2}}$,
$\hat p_2=\frac{-\hat\bx+\hat\by}{\sqrt{2}}$, we have
\begin{eqnarray}
M_{\frac{k(\hat\bx+\hat\by+\hat\bz)}{\sqrt{3}},t,p_1} &=&
N\frac{ik\tilde\gamma}{\sqrt{6}}[2\xi(\cos\theta-\cos\phi\sin\theta)\sin\theta
\sin\phi \nonumber \\
&&+ (\cos^2\theta-\cos^2\phi\sin^2\theta)(1-\zeta)] \label{M111txz}
\end{eqnarray}
and
\begin{eqnarray}
M_{{k(\hat\bx+\hat\by+\hat\bz)}/{\sqrt{3}},t,p_2} &=&
N\frac{ik\tilde\gamma}{\sqrt{6}}[\sin(2\theta)(\sin\phi-\cos\phi)\xi
\nonumber \\
&&-\sin^2\theta\cos(2\phi)(1-\zeta)] \label{M111txy}
\end{eqnarray}
\end{subequations}
respectively.

For the longitudinal wave, $\bu_{k(\hat\bx+\hat\by)/\sqrt{2},l}(\br) = N\frac{\hat\bx+\hat\by}{\sqrt{2}} e^{ik\frac{\hat\bx+\hat\by}{\sqrt{2}}\cdot\br}$, propagating in the $\langle110\rangle$ direction,
\begin{subequations}\label{M110}
\begin{eqnarray}
M_{k(\hat\bx+\hat\by)/\sqrt{2},l} &=& N
\frac{ik\tilde\gamma}{2}[2\sin(2\phi)\sin^2\theta\xi+(1+\cos^2\theta)\zeta
\nonumber \\
&& +\sin^2\theta] \label{M110l}
\end{eqnarray}
and for the two transversal waves in the same direction,
$\bu_{k(\hat\bx+\hat\by)/\sqrt{2},t,p^{\prime}_1}(\br)=N\hat p^{\prime}_1
\exp\left[ik\frac{\hat\bx+\hat\by}{\sqrt{2}}\cdot\br\right]$ and
$\bu_{k(\hat\bx+\hat\by)/\sqrt{2},t,p^{\prime}_2}(\br)=N\hat p^{\prime}_2
\exp\left[ik\frac{\hat\bx+\hat\by}{\sqrt{2}}\cdot\br\right]$,
with polarizations $\hat p^{\prime}_1=\hat\bz$,
$\hat p^{\prime}_2=\frac{-\hat\bx+\hat\by}{\sqrt{2}}$, we have
\begin{eqnarray}
M_{k(\hat\bx+\hat\by)/\sqrt{2},t,p^{\prime}_1} &=& N
\frac{ik\tilde\gamma\zeta}{\sqrt{2}}\sin(2\theta)(\sin\phi+\cos\phi)
\label{M110tz}
\end{eqnarray}
and
\begin{eqnarray}
M_{k(\hat\bx+\hat\by)/\sqrt{2},t,p^{\prime}_2} &=& N
\frac{ik\tilde\gamma}{2}\sin^2\theta\cos(2\phi)(\zeta-1),
\label{M110txy}
\end{eqnarray}
\end{subequations}
respectively.

Now we can calculate the phonon's scattering rates, by averaging
$\Gamma_{\bk\sigma}$ of Eq. (\ref{eqn_Gamma_bar}) over the
distribution of TLS parameters, $\Delta$, $\Lambda$, $\theta$
and $\phi$ and taking into account the scattering of phonons from and into the
mode $(\bk,\sigma)$. We assume that the parameters $\Delta$ and $\Lambda$ are
independent of the parameters $\theta$ and $\phi$, and their distribution
is the standard $P(\Delta,\Lambda)=P_0/\Lambda$, where $P_0$
is a constant
\cite{JLowTempPhys.7.351.1972.Philips,PhilMag.25.1.1972.Anderson}.
We change the variables $\Delta$ and $\Lambda$ into the variables
$\epsilon$ and $u\equiv\Lambda/\epsilon$, with the probability distribution
$P(\epsilon,u)=P_0/\left(u\sqrt{1-u^2}\right)$ and we assume that the
fraction of excited TLSs, of energy $\epsilon$, is thermal and corresponds to
a temperature $T$: $n^{(TLS)}_{\epsilon}=(1+e^{\epsilon/\kb T})^{-1}$.
The distribution over $\theta$ and $\phi$, say $f(\theta,\phi)$, is
unknown. Plugging all these quantities into the standard scattering rate
calculation, we get
\begin{eqnarray}
\tau^{-1}_{\bk\sigma} &=& \frac{P_0\tanh\left(\frac{\epsilon}{2\kbt}
\right)}{2\hbar}n_{\bk\sigma}\int_0^\pi\sin\theta\,d\theta\nonumber \\
&&\times\int_0^{2\pi}d\phi\cdot|M_{\bk\sigma}[\hat\bt(\theta,\phi)]|^2
f(\theta,\phi) \nonumber \\
&\equiv& \frac{P_0\tanh\left(\frac{\epsilon}{2\kbt}
\right)}{4\pi\hbar}n_{\bk\sigma}\langle|M_{\bk\sigma}(\hat\bt)|^2\rangle.
\label{av_def}
\end{eqnarray}
Now, if we would know $f(\theta,\phi)$, we could use the
expressions (\ref{M100}), (\ref{M111}), and (\ref{M110}) for $M$,
to calculate scattering rates of phonons propagating in the three
different directions.

Since, as we mentioned in the Introduction, we have no microscopic
model for $f(\theta,\phi)$, we shall assume that $f(\theta,\phi)$
is constant (i.e. TLSs are isotropically oriented) and comparing
our calculations with experimental results, we shall obtain
relations between the parameters $\zeta$ and $\xi$.  Using this
assumption and Eq. (\ref{M100l}), we get for the longitudinally
polarized phonon propagating in the $\langle100\rangle$ direction,
\begin{subequations}\label{av_100}
\begin{equation}
\tau^{-1}_{k\hat\bz,l}
= \frac{3+4\zeta+8\zeta^2}{15}\cdot\frac{2\pi P_0N^2nk^2\tilde\gamma^2}{\hbar}
\tanh\left(\frac{\epsilon}{2\kbt}\right) ,
\label{av_100l}
\end{equation}
while for the transversally polarized waves, both Eqs. (\ref{M100tx})
and (\ref{M100ty}) give the same result,
\begin{equation}
\tau^{-1}_{k\hat\bz,t}
= \frac{4\xi^2}{15}\cdot\frac{2\pi P_0N^2nk^2\tilde\gamma^2}{\hbar}
\tanh\left(\frac{\epsilon}{2\kbt}\right),
\label{av_100t}
\end{equation}
\end{subequations}
where we dropped also the obvious subscripts of the population number, $n$.

Similarly, in the direction $\langle111\rangle$ we get
\begin{subequations}\label{av_111}
\begin{eqnarray}
\tau^{-1}_{\frac{k(\hat\bx+\hat\by+\hat\bz)}{\sqrt{3}},l}
&=& \frac{5+20\zeta+20\zeta^2+16\xi^2}{45}\cdot\frac{2\pi P_0k^2N^2n
\tilde\gamma^2}{\hbar}\nonumber \\
&&\times\tanh\left(\frac{\epsilon}{2\kbt}\right) ,
\label{av_111l}
\end{eqnarray}
and
\begin{eqnarray}
\tau^{-1}_{\frac{k(\hat\bx+\hat\by+\hat\bz)}{\sqrt{3}},t}
&=& \frac{2[(1-\zeta)^2+2\xi^2]}{45}\cdot\frac{2\pi P_0k^2N^2n\tilde\gamma^2}
{\hbar}\nonumber \\
&& \times\tanh\left(\frac{\epsilon}{2\kbt}\right),
\label{av_111t}
\end{eqnarray}
\end{subequations}
where again, the two transversally polarized waves, Eqs. (\ref{M111txz})
and (\ref{M111txy}), give the same result.

Finally, for the phonons propagating along the $\langle110\rangle$
direction we obtain the average scattering rates,
\begin{subequations}\label{av_110}
\begin{eqnarray}
\tau^{-1}_{\frac{k(\hat\bx+\hat\by)}{\sqrt{2}},l}
&=& \frac{2+6\zeta+7\zeta^2+4\xi^2}{15}\cdot\frac{2\pi P_0k^2N^2n
\tilde\gamma^2}{\hbar}\nonumber \\
&& \times \tanh\left(\frac{\epsilon}{2\kbt}\right),
\label{av_110l}
\end{eqnarray}
for the longitudinal wave,
\begin{equation}
\tau^{-1}_{\frac{k(\hat\bx+\hat\by)}{\sqrt{2}},t,z}
= \frac{4\xi^2}{15}\cdot\frac{2\pi P_0k^2N^2n\tilde\gamma^2}{\hbar}
\tanh\left(\frac{\epsilon}{2\kbt}\right),
\label{av_110tz}
\end{equation}
for the transversal wave polarized in the $\hat p^{\prime}_1$ direction, and
\begin{equation}
\tau^{-1}_{\frac{k(\hat\bx+\hat\by)}{\sqrt{2}},t,p^{\prime}_2}
= \frac{(\zeta-1)^2}{15}\cdot\frac{2\pi
P_0k^2N^2n\tilde\gamma^2}{\hbar}
\tanh\left(\frac{\epsilon}{2\kbt}\right), \label{av_110txy}
\end{equation}
\end{subequations}
for the transversal wave polarized in the $\hat p^{\prime}_2$ direction.

In the STM formalism, with $\delta=2\gamma_\sigma S_\sigma$--$\sigma=l,t$--, 
the transition rates are \cite{PhysRevB75.064202.2007.Anghel}
\begin{equation}
\left(\tau^{(STM)}_{\bk,\sigma}\right)^{-1}
= \frac{2\pi P_0k^2N^2n\gamma^2_\sigma}{\hbar}
\tanh\left(\frac{\epsilon}{2\kbt}\right),
\label{av_STM}
\end{equation}
therefore the Eqs. (\ref{av_100})-(\ref{av_110}) give the expressions for the
$\gamma^2_\sigma$s for the phonons propagating in different crystalographic
directions. Notice that if we impose the condition for isotropy,
$\zeta+2\xi=1$, all equations (\ref{av_100})-(\ref{av_110}) reduce
to the isotropic expressions of Ref. \cite{PhysRevB75.064202.2007.Anghel}.

Karen Topp, Robert Pohl, and coworkers (see Ref. \cite{Topp:thesis} and
references therein) measured $\gamma$ in the crystalographic
directions $\langle100\rangle$ and $\langle111\rangle$ of the
cubic lattice of Ca stabilized Zirconium. They obtained a ratio
between $\gamma_t$ in the  $\langle111\rangle$ direction and
$\gamma_t$ in the $\langle100\rangle$ direction, of about 1.7.
Using this result and Eqs. (\ref{av_100t}) and (\ref{av_111t}), we
obtain a relation between $\zeta$ and $\xi$:
\begin{equation}
\left(\frac{1-\zeta}{\xi}\right)^2 = 18.4 . \label{rel_zeta_xi}
\end{equation}

Notice that the lower symmetry of the cubic lattice modifies the
relation $(1-\zeta)^2/\xi^2=4$, satisfied in an isotropic
medium, into the relation
(\ref{rel_zeta_xi}). More experimental data would enable one to
check this relation for other polarizations or propagation
directions and eventually even to calculate the elements of $[R]$.
If relation (\ref{rel_zeta_xi}) does not hold true for any
propagation direction and polarization of the elastic wave, then
the distribution of TLS orientations is not isotropic.

\section{Conclusions}

We applied the formalism introduced in Ref.
\cite{PhysRevB75.064202.2007.Anghel}
to describe the interaction of phonon modes (or elastic waves) with
the ensemble of two-level systems (TLS) in a disordered cubic crystal.
We showed that the interaction is anisotropic and--in the language
of the standard tunneling model--the coupling constants
$\gamma_l$ and $\gamma_t$ depend on the phonon propagation direction.
We focused our calculations on phonons propagating along the
crystalographic directions $\langle100\rangle$,
$\langle110\rangle$ and $\langle111\rangle$, for which we gave
explicit expressions for the coupling constants. Using the
experimental results of Topp, Pohl, and coworkers (see Ref.
\cite{Topp:thesis} and references therein) we compared the
$\gamma_t$s corresponding to the $\langle100\rangle$ and
$\langle111\rangle$ crystalographic directions and from here we
obtained a relation between the the two parameters of the model,
$\zeta$ and $\xi$, that describe the anisotropy of the
interaction. Nevertheless, more experimental results are needed
(at least $\gamma_t$ in one more direction or a $\gamma_l$) to
fully determine these parameters and make prediction about the
interaction of the TLS system with phonons propagating in any
direction. Having these predictions, one then could draw
conclusions about the isotropy of the TLS orientations in the
material.

\section*{Acknowledgements}

We are grateful to Prof. R. Pohl, Prof. K. A. Topp, and Prof. S. Sahling 
for very useful and motivating discussions. This work was partially 
supported by the NATO grant, EAP.RIG 982080.


\end{document}